\documentstyle[aps,prl,epsfig,float,twocolumn]{revtex}
\begin{document}
\twocolumn[\hsize\textwidth\columnwidth\hsize\csname
@twocolumnfalse\endcsname

\title{Non-ballistic spin field-effect transistor}

\author{John Schliemann, J. Carlos Egues\cite{carlosaddress}, and Daniel Loss}

\address{Department of Physics and Astronomy, University of Basel,
CH-4056 Basel, Switzerland}

\date{\today}

\maketitle

\begin{abstract}
We propose a spin field-effect transistor based on spin-orbit
(s-o) coupling of both the Rashba and the Dresselhaus types.
Differently from earlier proposals, spin transport through our
device is tolerant against spin-independent scattering processes.
Hence the requirement of strictly ballistic transport can be
relaxed. This follows from a unique interplay between the
Dresselhaus and the (gate-controlled) Rashba interactions; these
can be tuned to have equal strengths thus yielding
\textit{k-independent} eigenspinors even in two dimensions. We
discuss implementations with two-dimensional devices and quantum
wires. In the latter, our setup presents strictly
\textit{parabolic} dispersions which avoids complications arising
from anticrossings of different bands.
\end{abstract}
\vskip2pc]

In the recent years research in semiconductor physics has
been focused on the emerging field of spintronics. This key word
refers to the variety of efforts to use the electron spin rather
than, or in combination with, its charge for information
processing; or, even more ambitiously, quantum information
processing \cite{Awschalom02}. Among the most prominent
device proposals is the spin field-effect transistor (FET) due to
Datta and Das \cite{Datta90}. This proposal uses the Rashba
spin-orbit coupling to perform controlled rotations of spins of
electrons passing through an FET-typed device. This particular
spin-orbit interaction is due to the inversion-asymmetry of the
confining potential and is of the form \cite{Rashba60}
\begin{equation}
{\cal
H}_{R}=\frac{\alpha}{\hbar}\left(p_{x}\sigma^{y}-p_{y}\sigma^{x}\right),
\label{rashba}
\end{equation}
where $\vec p$ is the momentum of the electron confined in a
two-dimensional geometry, and $\vec\sigma$ the vector of Pauli
matrices. The coefficient $\alpha$ is tunable in strength by the
external gate of the FET. Due to the dependence on the momentum,
the Rashba spin-orbit coupling can be viewed as a wave
vector-dependent Zeeman field which can change drastically if the
electron is scattered into a different momentum state. Therefore,
such scattering events readily randomize the electron spin thus
limiting the range of operation of the Datta-Das spin-FET to the
regime of \textit{ballistic} transport where such processes do not
occur.

In the present work we propose a modified version of the spin-FET
in which the electrons are not only subject to spin-orbit
interaction of the Rashba but also of the Dresselhaus type
\cite{Dresselhaus55}. The latter is present in semiconductors
lacking bulk inversion symmetry. When restricted to a
two-dimensional semiconductor nanostruture with appropriate growth
geometry this coupling is of the form \cite{Dyakonov86,Bastard92}
\begin{equation}
{\cal H}_{D}=\frac{\beta}{\hbar}\left(p_{x}\sigma^{x}-p_{y}\sigma^{y}\right),
\label{dressel}
\end{equation}
where the coefficient $\beta$ is determined by the semiconductor
material and the geometry of the sample. Below we show that our
proposed device is robust against spin-independent scattering and
hence can also operate in a non-ballistic (or diffusive) regime.
This unique feature follows from the possibility of tuning the
Rashba (via proper gating) and the Dresselhaus terms so that they
have equal strengths $\alpha=\beta$. In this case, we show quite
generally below that the electron spinor is $k$-independent in two
dimensions -- even in the presence of (spin-independent)
scatterers.

\textit{Tuned Rashba and Dresselhaus terms.} Consider the
Hamiltonian ${\cal H}=\vec p^{2}/2m+V(\vec r)+{\cal H}_{R}+{\cal
H}_{D}$, where $m$ is the effective mass of the semiconductor and
$V(\vec r)$ an {\em arbitrary} scalar potential. For
$\alpha=\pm\beta$ the operator $\Sigma=(\sigma^{x}\pm\sigma^{y})/\sqrt{2}$
provides an additional conserved quantity, and a general
eigenstate of $\cal H$ and $\Sigma$ reads (for $\alpha=+\beta$)
\begin{equation}
\psi_{\pm}(\vec r)=\frac{1}{\sqrt{2}}
\left(
\begin{array}{c}
1 \\ \pm e^{i\pi/4}
\end{array}
\right) \varphi(\vec r)e^{\mp i\sqrt{2}\alpha m(x-y)/\hbar^{2}},
\label{eigenstate}
\end{equation}
where the function $\varphi(\vec r)$ fulfills the usual
spin-independent Schr\"odinger equation
\begin{equation}
\left(-\frac{\hbar^{2}}{2m}\nabla^{2} +V(\vec
r)\right)\varphi(\vec r)=
\left(\varepsilon+\frac{2\alpha^{2}m}{\hbar^{2}}\right)\varphi(\vec
r), \label{schroedinger}
\end{equation}
and $\varepsilon$ is the energy eigenvalue of the wave function
$\psi_{\pm}(\vec r)$ with $\Sigma=\pm 1$. Since
Eq.~(\ref{schroedinger}) is independent of the quantum number $\Sigma$
all eigenstates are generally twofold degenerate. Such two
degenerate states differing in $\Sigma$ are related by time reversal.
Note that the eigenvalue problem (\ref{schroedinger}) is invariant
under a formal time reversal, and the function $\varphi(\vec r)$
can be taken to be real. The potential $V(\vec r)$ can provide
further confinement of the quantum well into a quantum wire (see
below) or a quantum dot; it can also possibly describe
\textit{non-magnetic} scatterers due to imperfections or impurities

{\em{A robust two-dimensional spin-FET}}. We consider an FET
setup given by a
two-dimensional quantum well which is laterally contacted by two
spin-polarized contacts. In the vertical direction across the well,
an electric field tuning the Rashba coefficient $\alpha$ is provided by a
gate. The spin-polarized leads can be realized by ferromagnetic
metals, or by ferromagnetic semiconductors. The latter version
appears to be preferable with respect to the spin injection
properties of the interfaces \cite{Awschalom02}.

Within the two-dimensional channel the Hamiltonian is ${\cal
H}=\vec p^{2}/2m+{\cal H}_{R}+{\cal H}_{D}$ whose eigenstates are
\begin{equation}
\psi_{\vec k}^{\pm}(\vec r)=\frac{1}{\sqrt{2}}\left(
\begin{array}{c}
1 \\
\pm e^{i\phi(\vec k)}
\end{array}\right)\frac{e^{i\vec k\vec r}}{2\pi},
\label{eigstate}
\end{equation}
with
$\phi(\vec k)=\arg(-\alpha k_{y}+\beta k_{x}+i(\alpha k_{x}-\beta k_{y}))$
and eigenenergies
\begin{equation}
\varepsilon(\vec k)=\frac{\hbar^{2}\vec k^{2}}{2m}
\pm\sqrt{\left(\alpha k_{y}-\beta k_{x}\right)^{2} +\left(\alpha
k_{x}-\beta k_{y}\right)^{2}}. \label{eigerg}
\end{equation}
For general $\alpha$ and $\beta$ the spinor in the eigenstates
(\ref{eigstate}) depends via $\phi(\vec k)$ on the wave vector,
and the dispersion (\ref{eigerg}) is non-parabolic. However, as
described above, the case $\alpha=\beta$ is particular. Here the
spin state of the wave functions is independent of the wave
vector, and the dispersion is perfectly parabolic. The first
observation is crucial for our device proposal.

\textit{Device operation.} The ``off state'' of our transistor
corresponds to a gate bias such that the Rashba and the
Dresselhaus coupling strengths are unequal, i.e., $\alpha \neq
\beta$. In this case, the spinor of an injected electron is $k$
dependent, Eq. (\ref{eigstate}), and hence becomes randomized due
to momentum scattering.  For strong enough spin relaxation, the
drain current is that of an unpolarized beam. The predominant
spin-dephasing mechanism in spin transistors is that of the
Dyakonov-Perel type \cite{Dyakonov71,Pikus84}, 
due to the largest of the two s-o
terms \cite{Bournel98}. The ``on state'' of our device operates with
a gate bias for which $\alpha=\beta$. Here an injected electron
with arbitrary momentum and in one of the $k$-independent spin
states $(1,\pm\exp(i\pi/4))$ traverses the transistor channel with
its spin state unchanged. The current at the drain would be the
same as the injected one, assuming that the ferromagnetic source
and drain have parallel polarizations. Note that spin-independent
scattering events (provided by lattice imperfections, phonons, and
non-magnetic impurities) cannot change the spin state of the
traversing electron. Moreover, as it will become clear from the
discussion below, further device setups can be
thought of as switching between the two points $\alpha=\pm\beta$
and/or using different 
combinations of magnetic polarizations in the contacts.

\begin{figure}
\centerline{\epsfig{file=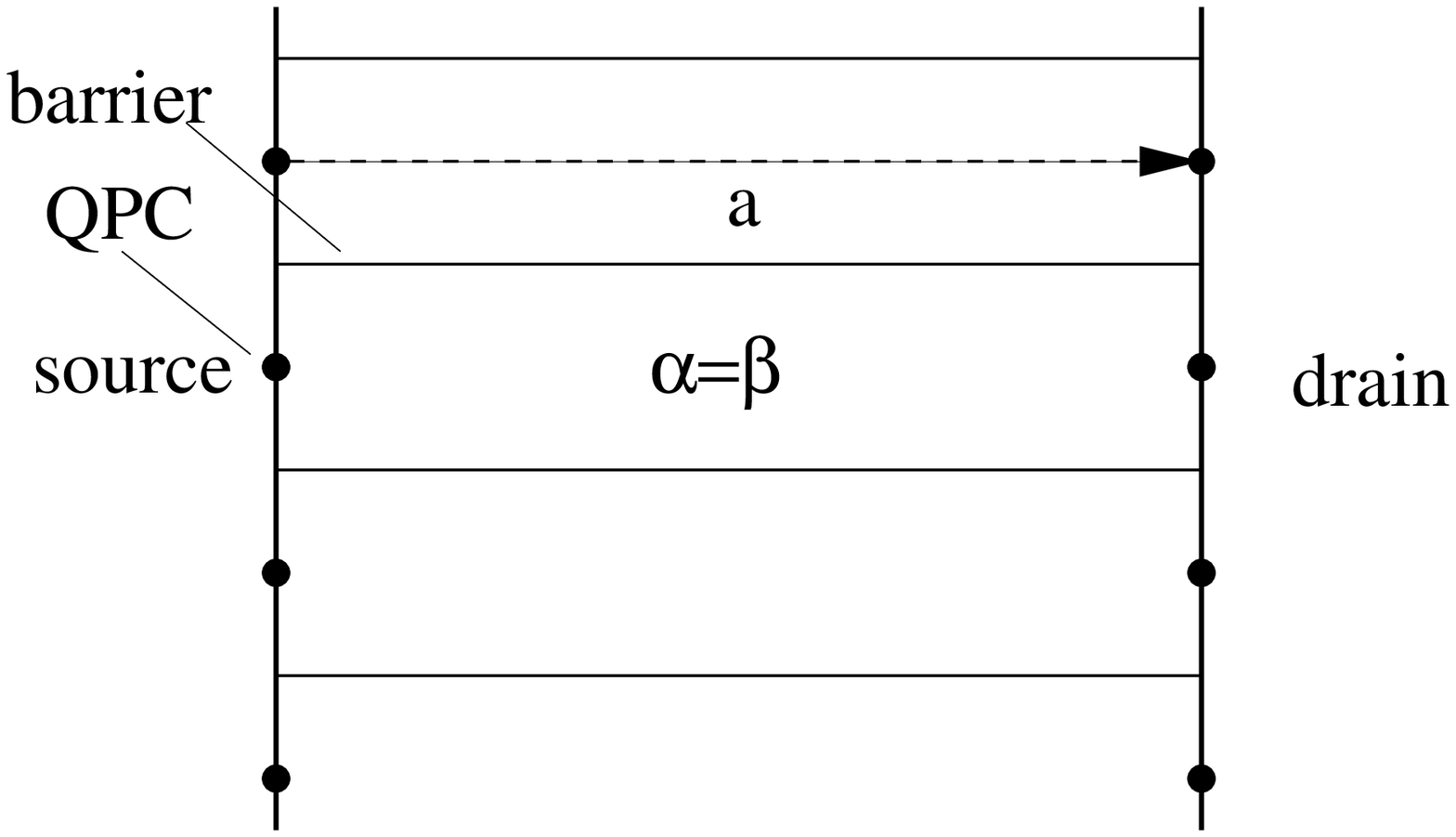, width=8cm}} \caption{Schematic
of the spin-FET setup using quantum point contacts (QPC) to source and drain.
The pairs of QPCs are separated by barriers to avoid crosstalk.} 
\label{fig1}
\end{figure}

\textit{Absence of spin relaxation.} The Elliot-Yafet
spin-flip mechanism is completely suppressed for $\alpha=\beta$
(``on state'' of our device) since the spinor is $k$ independent
in this case. In addition, the Rashba-Dresselhaus rotation axis is
fixed for equal couplings and hence no Dyakonov-Perel spin relaxation is
operative either. This can be seen from the general stationary
solution (\ref{eigenstate}): particles injected into the device
with spin components in one of the eigenspinor states
$(1,\pm\exp(i\pi/4))$ do not get altered at all (up to an
unimportant global phase). Moreover, inspection of Eq.~(\ref{eigenstate})
shows that particles injected in a
general spin state do not undergo a randomization of their spin
but a controlled rotation around the $(1,1,0)$ axis by an angle
$\eta$ given by $\eta=2\sqrt{2}\alpha m(a_{x}-a_{y})/\hbar^{2}$,
where $\vec a$ connects the locations where the particles are injected
and detected, respectively. Therefore, if the 
locations of injection and detection are defined with
sufficient precision, an electron injected in a general spin state
will not suffer a randomization of it spin. However, uncertainties
in those locations will translate to an uncertainty in the rotation angle.
A way to avoid this problem is to inject electrons in the eigenspinor
states $(1,\pm\exp(i\pi/4))$ (as discussed above)
where the rotation has only a trivial effect,
or to inject and detect electrons in a general spin state through quantum point
contacts leading to well-defined distance vector $\vec a$. 
To enable a higher efficiency of the device, arrays of such quantum point
contact pairs, separated by barriers, can be used in parallel as shown
schematically in figure \ref{fig1}.

A similar
finding was obtained numerically by Kiselev and Kim
\cite{Kiselev00} who studied an effective spin model of the form
$\tilde{\cal H}={\cal H_{R}}+{\cal H_{D}}$, where the
momentum $\vec p(t)=m\dot{\vec r}$ is a classical variable (not an
operator) whose dependence on time $t$ is generated by a Markovian
process. The general time evolution operator reads ${\cal
U}(t)={\cal T}\exp(-i\int_{0}^{t}dt'\tilde{\cal H}[\vec p(t')]/\hbar)$,
where $\cal T$ denotes the time-ordering symbol. For
$\alpha=\beta$ the time ordering becomes trivial, and ${\cal
U}(t)$ reads, up to a global phase,
\begin{equation}
{\cal U}(t)={\bf 1}\cos\left(\frac{\eta(\vec a)}{2}\right)
-i\frac{\sigma^{x}+\sigma^{y}}{\sqrt{2}}
\sin\left(\frac{\eta(\vec a)}{2}\right)\,,
\label{evolution}
\end{equation}
with, as above, $\eta(\vec a)=2\sqrt{2}\alpha
m(a_{x}-a_{y})/\hbar^{2}$ and $\vec a=\vec r(t)-\vec r(0)$. Note
that this finding is independent of whether or not the energy
$\tilde{\cal H}[\vec p(t)]$ is conserved along the path $\vec p(t)$, as
it was assumed in Ref.~\cite{Kiselev00}. Thus, also this
simplified effective spin model (with the orbital degrees of
freedom treated classically) leads to the same controlled spin
rotation as the full quantum mechanical solution
(\ref{eigenstate}). The above findings are in contrast with
earlier assertions where a randomization of the spin was predicted
to occur even for $\alpha=\beta$ \cite{Pikus95}, or at least for
$\alpha=\beta$ and a general spin state of the injected electron
differing from $(1,\pm\exp(i\pi/4))$ \cite{Averkiev99}. 
These conclusions are due to the weak-coupling 
treatment performed in \cite{Pikus95,Averkiev99} and become invalid 
in the presence of the additional conserved quantity $\Sigma$
arising at $\alpha=\beta$.

A quantitative description of the transport in the ``on state'' of
our device should include possible spin-independent scatterers in
the potential $V(\vec r)$ of Eq.~(\ref{schroedinger}). This
equation describes the \textit{orbital} part of the
single-particle eigenstates whose spin part is independent of the
momentum for $\alpha=\beta$. Solutions to this equation can be
matched with wave functions in the leads according to the
appropriate boundary conditions in the presence of spin-orbit
coupling \cite{Molenkamp01,Matsuyama02}. Transmission
coefficients can then be determined from the stationary solutions.

\textit{Ballistic regime.} In the strictly ballistic case ($V(\vec
r)=0$) and for source and drain with parallel polarizations chosen
along either of the spinor directions $(1,\pm\exp(i\pi/4))$, we
find the transmission amplitude
\begin{equation}
T^{\pm}= \frac{e^{-ika\pm i\sqrt{2}\alpha m_{2}a/\hbar^{2}}
\left(4q\frac{m_{2}}{m_{1}}k\right)}
{\left(\frac{m_{2}}{m_{1}}k+q\right)^{2}e^{-iq a}
-\left(\frac{m_{2}}{m_{1}}k-q\right)^{2}e^{iq a}},
\label{eq-trans}
\end{equation}
for an electron injected at energy $\varepsilon$ and wave vector
$\vec k=k\vec e_{y}$. In (\ref{eq-trans}) $a$ is the length of
s-o active region, $m_{1}$, $m_{2}$ are the band masses in
the contacts and the two-dimensional channel, respectively, and
$q=\sqrt{2m_{2}(\varepsilon-V_{0})/\hbar^{2}}$ with $V_{0}$ being
a possible band offset between the contacts and the FET channel (also
including a contribution from s-o coupling). 
From the above expression one can find the conductance using Landauer's
fromula. Concerning the phase of transmitted electrons one finds
$T^{+}/T^{-}=\exp(i2\sqrt{2}\alpha m_{2}a/\hbar^{2})$. As discussed above,
this phase factor is also obvious from the 
general form of eigenstates (\ref{eigenstate}) and corresponds to a
controlled rotation of the spin of the injected particle around the
(1,1,0) direction. The rotation angle is, up to a factor of $\sqrt{2}$, due to
the presence of both Rashba and Dresselhaus coupling, the same as
the one in the original proposal by Datta and Das \cite{Datta90}.
Note, however, that here the spin part of the wave
functions is independent of the wave vector. Moreover, according
to the general form of eigenstates given by
Eq.~(\ref{eigenstate}), the same phase factor occurs if
spin-independent scatterers encompassed in the potential $V(\vec
r)$ are included. Therefore, as discussed above,  the range of operation 
of our device is not limited to the strictly ballistic regime.

\textit{Magnitudes of $\alpha$ and $\beta$.} The largest values
for $\alpha$ observed in III-V semiconductors are of the order of
a few $0.1{\rm eV\AA}$
\cite{Nitta97,Engels97,Heida98,Hu99,Grundler00,Sato01}. An
estimate for the Dresselhaus coefficient in a confined geometry is
obtained from $\beta=\gamma\langle k_{z}^{2}\rangle$, where
$\langle k_{z}^{2}\rangle$ is the expectation value of the square
wave vector component in the growth direction. A typical value for
the coefficient $\gamma$ is $\gamma\approx 25{\rm eV\AA}^{3}$
\cite{Lommer88,Jusserand92,Jusserand95}. For an infinite well with
width $w$ we find $\langle k_{z}^{2}\rangle=(\pi/w)^{2}$, which
yields $\beta\approx 0.09{\rm eV\AA}$ for $w=50{\rm \AA}$. Hence
there should be no principle difficulty to achieve the situation
$\alpha=\beta$ even in comparatively narrow wells.
Note also that small deviations from the case $\alpha=\beta$, i.e.
$\alpha=\beta+\delta$ with $|\delta/\alpha|\ll 1$ lead (using Fermi's
golden rule) to spin scattering rates which are quadratic in $\delta$.
Thus, spin dephasing due to spin-orbit coupling is completely suppressed 
in first order in $\delta$. This is in accordance with the results of
Ref.~\cite{Kiselev00} studying an effective time-dependent Hamiltonian
where the inverse dephasing time has a minimum equal to zero at
$\alpha=\beta$ and is differentiable around this point.

{\em{Quantum wire with spin-orbit coupled bands.}} We now consider
a quantum wire formed by an additional confining potential $V(x)$
along the $x$-direction. For $\alpha=\beta$ single-particle wave
functions are of the form (\ref{eigenstate}) with $\varphi_n(\vec
r)=\chi_n(x)\exp(i(k\mp\sqrt{2}\alpha m/\hbar^{2})y)$, where
$\chi_{n}(x)$ obeys the usual Schr\"odinger equation for the
transverse variable $x$ with quantized eigenvalues
$\tilde\varepsilon_n$, $n$ labels the energy levels. The
single-particle eigenenergies are then
$\varepsilon_n^{\pm}(k)=\tilde\varepsilon_n+(\hbar^{2}/2m)
(k\mp\sqrt{2}\alpha m/\hbar^{2})^{2}-2\alpha^{2}m/\hbar^{2}$. Note
that, similarly to the two-dimensional case discussed earlier, the
wire energy dispersions here are also parabolic -- for
\textit{any} strength of the $\alpha=\beta$ coupling. In addition,
as we discuss below, there are no avoided crossings in the energy
dispersions. These results are significantly different from the
usual case of a quantum wire with only the Rashba s-o interaction
\cite{Moroz99,Mireles01,Governale02}; there the bands are highly
non-parabolic and anti-cross for strong Rashba couplings.

Figure 2 illustrates the wire dispersions for a two-band model.
Note in Fig. 2(a) the features mentioned above for the case with
tuned couplings: parabolic dispersions with no anticrossings. For
differing coupling constants $\alpha\neq\beta$, the bands are
non-parabolic and display avoided crossings. The contrasting
features of the $\alpha=\beta$ and $\alpha\neq\beta$ case 
are crucial for spin injection across a no-s-o/s-o active
interface.

\textit{Spin injection in quasi-1D channels.} Strong Rashba s-o
interaction can greatly affect the spin conductance of wires
\cite{Mireles01} and even suppress spin injection
\cite{Governale02}. The wires we consider here, with tuned s-o
couplings $\alpha=\beta$, should not present any obstacle to spin
injection since the bands are parabolic with no avoided crossings.
The problem with tuned couplings is similar to that of a quantum
wire with \textit{uncoupled} s-o bands, where spin injection is
always possible in the ballistic regime \cite{Datta90}.

We can also consider a spin-FET
setup with a wire as the connecting channel between the source and
the drain. Since the spin part of the wire eigenstates are
wave-vector dependent for $\alpha\neq\beta$, this quasi-1D spin FET
operates similarly to the non-ballistic two-dimensional one
discussed earlier. That is, elastic and/or inelastic scattering
processes changing the wave vector also randomize the spin state
of transmitted electrons (``off state'') for wires with many
bands. These effects are absent for $\alpha=\beta$ and the spin
state is preserved(``on state'').

\textit{Inflences beyond the effective Hamiltonian.} 
Our analysis assumes that the effects of
spin-orbit interaction are entirely described by the contributions
(\ref{rashba}) and (\ref{dressel}) to the Hamiltonian. In a
realistic semiconductor system there are additional 
corrections to these dominant terms. 
It is instructive to consider the influence of possible
nonparabolicity in the band structure described by a contribution to the 
Hamiltonian of higher order in the momentum.
For instance, as it was argued theoretically \cite{Lommer88} and confirmed 
experimentally \cite{Hu99}, particularly large values of $\alpha$ are 
typically accompanied by a sizeable quartic nonparabolicity
of the form $(\vec p^{2})^{2}$.
This is due to the similar dependence of both
terms on the band gap. However, also in this case $\Sigma$ is still
a conserved quantity at $\alpha=\beta$, and the Hamiltonian is
invariant under time reversal if only even powers of the momentum occur.
In particular, if 
$(1,\exp(i\pi/4)\Phi(\vec r)$ solves the stationary Schr\"odinger equation,
so does the orthogonal state 
$(1,-\exp(i\pi/4)\Phi^{*}(\vec r)$ with the same
energy, thus leading to the same general 
degeneracy pattern as in the parabolic case.
Moreover, transmission amplitudes for
such eigenspinor states are as before related by complex conjugation. 
Thus, for injection into eigenspinor states $(1,\pm\exp(i\pi/4)$
the device operation is completely unaltered. For injection into
linear combinations of them the same controlled rotation around
the $(1,1,0)$ axis occurs. The rotation angle, however, is more difficult
to determine since an elegant transformation as described in
Eqs.~(\ref{eigenstate}),(\ref{schroedinger}) does not seem to exist.
Note that the result for the spin evolution operator (\ref{evolution})
obtained within the classical approximation remains the same
if additional spin-independent terms are included, and the rotation angle is
again given by the distance $\vec a $. 
In summary, in situations
where a sizeable quartic term is present and the classical
approximation to the orbital motion appears problematic, 
injection in directions close to the eigenspinor directions 
is favorable in order to ensure good device operation.
Thus, even with such
corrections like nonparabolicity being included,
our spin-transistor proposal -- which
benefits from a unique ``cancellation'' of the Rashba and the
Dresselhaus terms for tuned couplings -- should provide a
substantial increase in performance and stability of a spin-FET
device as compared with the original proposal \cite{Datta90}. We
stress that this cancellation occurs for both signs in the
relation $\alpha=\pm\beta$. Therefore further devices can be
envisioned switching between these two points and/or using different 
combinations of magnetic polarizations in the contacts.

This work was supported by NCCR Nanoscience, the Swiss NSF,
DARPA, and ARO.

\begin{figure}
\centerline{\epsfig{file=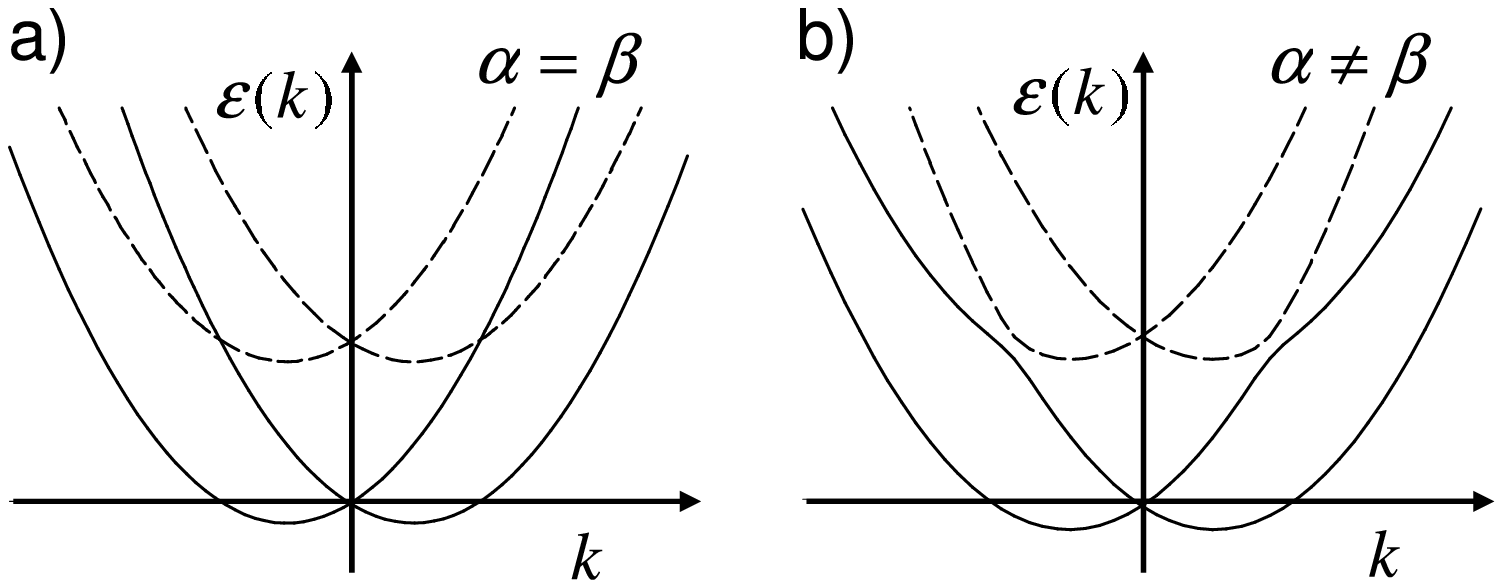, width=8cm}} \caption{Quantum
wire dispersions $\varepsilon(k)$
in the presence of both Rashba and Dresselhaus
s-o interactions. For equal s-o strengths $\alpha=\beta$
(a) the dispersions are
parabolic with no anti crossings. For differing coupling strengths
$\alpha\neq\beta$
(b) the bands are non-parabolic and avoided crossings occur. }
\label{fig2}
\end{figure}

\end{document}